\documentclass[reprint,aps,amsmath,amssymb]{revtex4-2}

\usepackage{graphicx}
\usepackage{dcolumn}
\usepackage{bm}
\usepackage[mathlines]{lineno}
\usepackage{lipsum}
\usepackage{xcolor}
\usepackage{svg}
\usepackage{siunitx}
\begin{document}

\preprint{APS/123-QED}

\title{Phenomenological Growth Regimes in Liquid-Precursor CVD of MoS$_2$ on Functional Substrates}

\author{$^1$Osamah Kharsah}
\author{$^2$Kilian Mouchel}
\author{$^1$Yossarian Liebsch}
\author{$^1$Cathy Sulaiman}
\author{$^1$Joel Verlande}
\author{$^1$Anke Hierzenberger}
\author{$^1$Abdallah Alghazali}
\author{$^2$Clara Grygiel}
\author{$^2$Stéphane Guillous}
\author{$^2$Henning Lebius}
\author{$^1$Marika Schleberger}

\affiliation{$^1$Faculty of Physics and CENIDE, University of Duisburg-Essen, 47057 Duisburg, Germany}

\affiliation{$^2$Université Caen Normandie, CEA, ENSICAEN, CNRS, Normandie Univ, CIMAP UMR6252, Caen, France}

\date{\today}

\begin{abstract}
The integration of two-dimensional transition-metal dichalcogenides (TMDCs) onto functional substrates remains constrained by stochastic vapor-phase growth dynamics. Here, we show that liquid-phase precursor chemical vapor deposition (CVD) of MoS$_2$ introduces growth conditions that are consistent with a substrate-influenced reaction-diffusion process. By utilizing pre-growth spin-coated MoO$_3$ intermediates across a diverse crystalline library (sapphire, SrTiO$_3$, rutile TiO$_2$, MgO, and 6H-SiC), we find that substrate-dependent variations in precursor wetting, surface chemistry, and inferred mass-transport constraints correlate with distinct growth morphologies. 
These substrate-dependent growth regimes are interpreted in terms of reduced effective lateral growth length on SrTiO$_3$, possible precursor anchoring on TiO$_2$, likely chemical surface restructuring on MgO, and possible step-edge growth on SiC. Raman and photoluminescence spectroscopy reveal substrate-dependent variations in vibrational and optical response that correlate with differences in strain, charge environment, and dielectric screening. Ultimately, this work highlights a substrate-dependent reaction-diffusion framework as a potentially useful route for tuning the structural and optical properties of large-area 2D materials.
\end{abstract}

\maketitle

\section{Introduction}
\label{sec:introduction}

\begin{table*}[ht]
\caption{Key physical and surface properties of the investigated substrate library. Variations in in-plane lattice constant ($a$), dielectric screening ($\epsilon_r$), and experimentally determined wetting properties are used here as phenomenological descriptors for the initial precursor distribution and substrate--precursor interaction. The measured macroscopic wettability (water contact angle, $\theta$) is complemented by the total surface free energy ($\gamma_{tot}$), estimated from water and diiodomethane contact angles and separated into dispersive ($\gamma^d$) and polar ($\gamma^p$) components. We note that these values were measured on pristine substrates and therefore serve as proxies for the initial wetting tendency rather than direct measurements of the high-temperature growth interface.}
\label{tab:substrate_properties}
\renewcommand{\arraystretch}{1.9}
\begin{tabular}{l c c c c c c}
\hline\hline
\textbf{Substrate}  &   \textbf{$a$ (\AA)}    &    \textbf{$\epsilon_r$}   &   \textbf{ $\theta$ ($^\circ$)} & \textbf{$\gamma_{tot}$ (mJ/m$^2$)} & \textbf{$\gamma^d$ (mJ/m$^2$)} & \textbf{$\gamma^p$ (mJ/m$^2$)} \\
\hline
Al$_2$O$_3$~(0001) & 4.76 \cite{Lee.1985} & $\sim 10$ \cite{Smink.2024} & 61.41 & 50.17 & 37.81 & 12.36 \\
SrTiO$_3$~(100)  & 3.90 \cite{Solokha.2019} & $\sim 300$ \cite{Tumarkin.2024} & 23.46 & 69.43 & 33.71 & 35.72 \\
Rutile TiO$_2$~(001)  & 4.59 \cite{Cromer.1955} & $\sim 86$  \cite{Parker.1961} & 19.79 & 70.54 & 32.59 & 37.96 \\
MgO~(100)  & 4.21 \cite{Hazen.1976} & $\sim 9.8$ \cite{Fontanella.1974} & 52.15 & 52.54 & 32.11 & 20.43 \\
SiC~(0001) & 3.08 \cite{INSPECInformationservice.1995} & $\sim 9.7$ \cite{INSPECInformationservice.1995} & 66.33 & 45.43 & 34.53 & 10.90 \\
\hline\hline
\end{tabular}
\end{table*}

Integrating two-dimensional (2D) transition-metal dichalcogenides (TMDCs) such as MoS$_2$ onto functional substrates is central to advanced optoelectronics and nanoelectronics, owing to their fast excitonic dynamics~\cite{Sim.2013}, tunable band gaps~\cite{Xi.2014}, and mechanical stability~\cite{Hou.2024}. In particular, substrate coupling enables tailored, application-specific functionalities. For example, the high dielectric constant of SrTiO$_3$ enables control over trionic luminescence~\cite{Chen.2018, Huang.2021,Haastrup.2023}, TiO$_2$ facilitates efficient interfacial charge transfer through Type-II band alignment for optoelectronic devices~\cite{Kharsah.2026b, Singh.2024, Liu.2019,Paul.2018}, and the polar covalent lattice of SiC supports heteroepitaxy and ultrafast optoelectronics~\cite{Xiao.2020, Ding.2023}.

Achieving such functionality requires clean, reproducible, and well-defined interfaces \cite{Kharsah.2026}. However, transfer-based integration is difficult to scale up and introduces water and polymer contamination, wrinkling, and trapped residues that degrade device performance~\cite{Huang.2026,Shen.2021}. Direct growth via chemical vapor deposition (CVD) is therefore preferred~\cite{Dong.2016,Zheng.2023}, yet conventional solid-precursor CVD is often affected by spatially variable vapor transport~\cite{GovindRajan.2016}. Consequently, nucleation and growth can be dominated by local precursor flux, which can obscure systematic substrate-dependent effects~\cite{Tang.2020,Seravalli.2021}.

\begin{figure*}[t]
\includegraphics[width=1\textwidth]{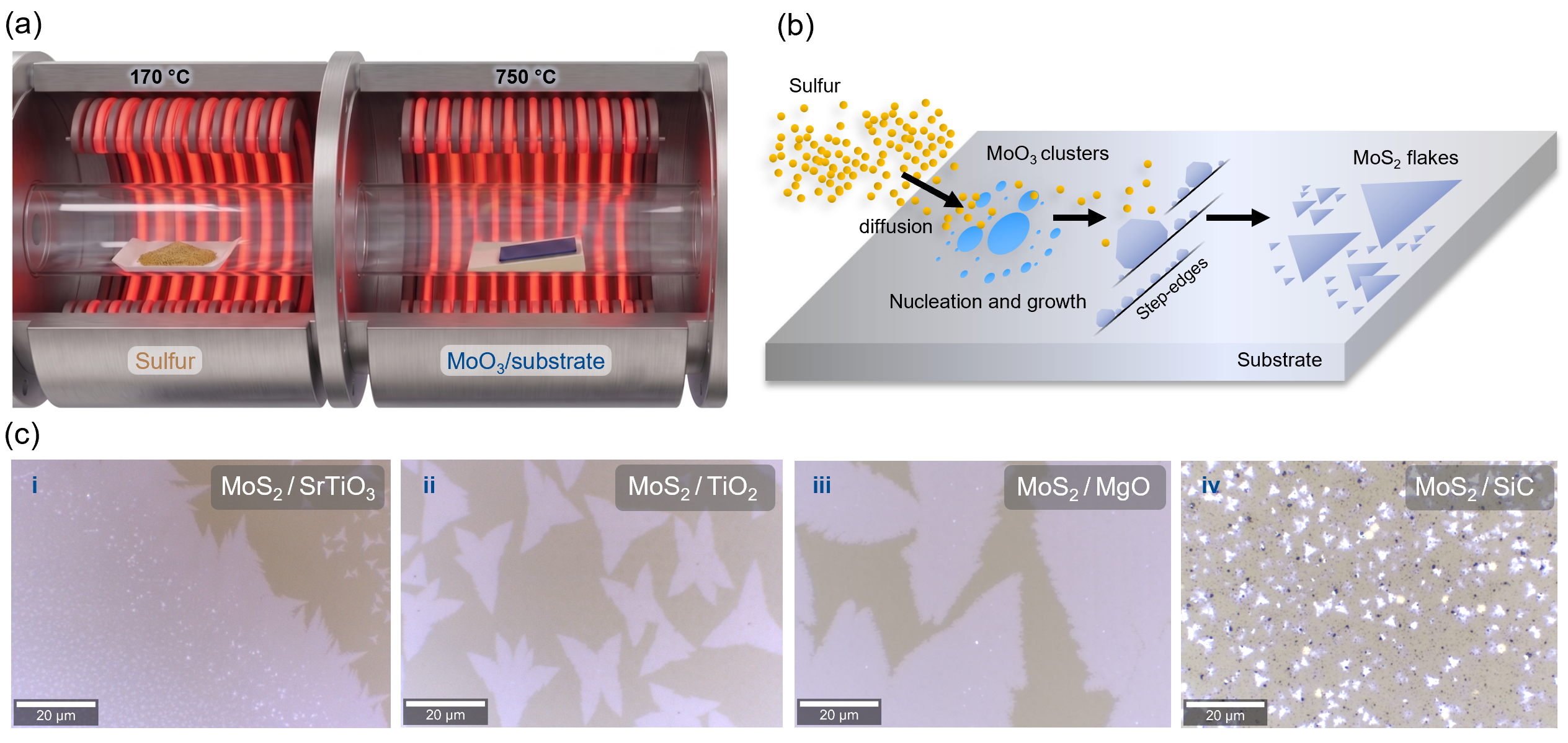}
\caption{
\textbf{(a)} Schematic of the CVD furnace used for MoS$_2$ growth. Sulfur is evaporated at 170~°C and transported via an Ar carrier gas to the pre-deposited MoO$_3$ on the growth substrate, where the MoS$_2$ growth process occurs at 750~°C. The schematic was generated with the assistance of Gemini (Google) and subsequently reviewed and corrected by the authors to represent the experimental setup.
\textbf{(b)} Schematic representation of the proposed liquid-precursor growth pathway, in which pre-deposited MoO$_3$-rich regions react with sulfur-containing species during sulfurization, leading to MoS$_2$ nucleation and lateral domain growth.
\textbf{(c)} Optical microscope images of MoS$_2$ grown on (i) SrTiO$_3$, (ii) TiO$_2$, (iii) MgO, and (iv) SiC.
}
\label{Fig:1}
\end{figure*}

Recently, liquid-phase precursor approaches have emerged as scalable alternatives, in which molybdenum oxide (MoO$_3$) is spin-coated and subsequently sulfurized~\cite{Esposito.2023,An.2022}. This is visualized in Figure~\ref{Fig:1}a, where sulfur is evaporated and transported via an Ar carrier gas to the pre-deposited MoO$_3$ on the growth substrate, located in a different heating zone. 
In this configuration, MoO$_3$ is initially localized on the substrate while the sulfur supply may be subject to gas-phase diffusion and surface-transport constraints, making the subsequent sulfurization increasingly sensitive to substrate-dependent precursor distribution and surface processes (illustrated in Figure~\ref{Fig:1}b). As a result, substrate physical (e.g., step-edges, wettability) and chemical (e.g., hydroxylation) properties likely influence local activation energies and sulfur diffusion lengths governing nucleation and crystallization~\cite{Kandybka.2024,Kalt.2023}. While conventional vapor-phase CVD has been successfully demonstrated on various functional substrates \cite{Haastrup.2023,Kropp.2020,Zheng.2023,Ding.2023}, liquid-precursor CVD has mainly focused on relatively inert substrates such as sapphire and SiO$_2$. Consequently, the physical framework governing how these substrate-precursor interactions influence growth dynamics on functional crystalline substrates has not been explored.

In this work, we investigate liquid-precursor CVD growth of MoS$_2$ on Al$_2$O$_3$~(0001), SrTiO$_3$ (100), rutile TiO$_2$~(001), MgO (100), and 6H-SiC~(0001). These substrates were selected to provide a range of surface chemistries, wetting properties, dielectric responses, and surface symmetries relevant to functional optoelectronic platforms. Rather than isolating any one of these properties, the substrate library allows us to examine how their combined influence correlates with precursor deposition and the resulting MoS$_2$ morphology. All substrates were processed under the same nominal deposition, calcination, and sulfurization conditions.

Optical microscopy and atomic force microscopy reveal pronounced substrate-dependent differences in MoS$_2$ coverage, domain size, domain shape, and multilayer fraction. We interpret these observations using a phenomenological reaction--diffusion framework that considers the initial precursor distribution, local conversion kinetics, effective mass transport, and possible substrate surface modification. Raman and photoluminescence spectroscopy further show that the resulting morphologies are accompanied by differences in the vibrational and optical response of MoS$_2$, consistent with combined contributions from strain, charge environment, structural inhomogeneity, and dielectric screening. These results demonstrate that the growth substrate is an active parameter in liquid-precursor CVD and must be considered in the direct integration of MoS$_2$ with functional material platforms.

\section{Results and Discussion}

\subsection*{Growth Configuration and Substrate Library}

In liquid-precursor CVD (see Section~\ref{methods}), the molybdenum precursor is pre-deposited on the growth substrate, while sulfur is supplied from the vapor phase during the high-temperature sulfurization step. Figure~\ref{Fig:1}a and b schematically illustrate the growth process. Because the Mo-containing precursor is deposited onto the growth substrate via an aqueous solution before sulfurization, the initial precursor distribution is expected to be sensitive to the substrate's wetting and surface chemical properties. The sulfur supply, in contrast, remains governed by gas-phase diffusion and subsequent surface transport to Mo-rich reaction sites. To examine how substrate properties influence this growth process, we performed liquid-precursor CVD on a crystalline substrate library consisting of Al$_2$O$_3$, SrTiO$_3$, rutile TiO$_2$, MgO, and SiC. We selected these substrates to span distinct surface chemistries, wetting properties, dielectric environments, and surface symmetries. The surfaces of SrTiO$_3$ (100), MgO (100), and rutile TiO$_2$~(001) have square or tetragonal surface nets \cite{Erdman.2002,Perry.1997,Diebold.2003}, whereas Al$_2$O$_3$~(0001) and 6H-SiC~(0001) have hexagonal surface symmetry \cite{Soares.2002,Starke.1997}. The relevant lattice constants ($a$), dielectric constants ($\epsilon_r$), as well as the measured water contact angles ($\theta$) and surface free energy ($\gamma_{tot}$) components -dispersive ($\gamma^d$) and polar ($\gamma^p$)- are summarized in Table~\ref{tab:substrate_properties}. 

These measurements nevertheless provide only indirect descriptors of precursor deposition. The actual precursor contains ammonium heptamolybdate and OptiPrep and is deposited after a sodium-cholate treatment; its wetting behavior may therefore differ from that of pure water on an untreated substrate. Furthermore, the room-temperature measurements do not represent the surface after calcination or during high-temperature sulfurization. Thus, we use the contact angle and surface-energy values only as comparative proxies for the initial wetting tendency, not as direct measures of the growth kinetics. This library enables comparison of oxide and carbide surfaces, hydration-stable and hydration-prone materials, weakly and strongly wetting surfaces, and substrates with markedly different dielectric constants. These differences may affect precursor deposition, nucleation, and edge-growth kinetics. Surface symmetry and lattice periodicity may also influence domain orientation. However, the MoS$_2$ lattice constant is not, by itself, sufficient to establish lattice matching. Such an analysis would require an experimentally determined in-plane orientation relationship or a defined coincidence lattice. We therefore discuss crystallographic templating only as a possible contribution and do not claim epitaxial growth from domain shape alone. 

\subsection*{Substrate-dependent Growth Behavior}
Optical microscopy reveals that the growth of MoS$_2$ varies significantly across the substrate library under otherwise identical growth conditions. Representative optical micrographs are shown in Figure~\ref{Fig:1}c. These images establish that liquid-precursor CVD does not produce a universal growth morphology on crystalline substrates. Instead, the final MoS$_2$ coverage, domain size, domain shape, and multilayer fraction appear to depend on the underlying surface. As a result, multiple growth regimes arise. 

The first growth regime is observed on SrTiO$_3$. Figure~\ref{Fig:1}c i reveals a high MoS$_2$ nucleation density, where numerous small domains ($\approx 5~\mu$m) coalesce into a nearly continuous film (Figure S1a). The extracted monolayer and multilayer coverages are approximately 66\% and 23\%, respectively. The dense distribution of nuclei and serrated coalescence fronts are consistent with efficient wetting of the aqueous precursor on hydrophilic SrTiO$_3$ \cite{Kawasaki.2017}, which may stabilize a large number of spatially distributed Mo-rich sites. The limited lateral expansion is consistent with a reduced effective growth length, although the present data do not distinguish between reduced surface mobility and competition between neighboring nuclei.

In contrast to SrTiO$_3$, MoS$_2$ nucleation density on TiO$_2$ is reduced, resulting in larger domains with star-like geometries (Figure~\ref{Fig:1}c ii). On the scale of the substrate, domains can exceed 500~$\mu$m (Figure S1b), with extracted monolayer and multilayer coverages of 44.3\% and 23.7\%, respectively.
This is surprising considering that the wettability and polar surface energy components of pristine TiO$_2$ and SrTiO$_3$ are similar ($\theta_{TiO_2} \approx 19.8^\circ$ vs. $\theta_{SrTiO_3} \approx 23.5^\circ$, and $\gamma^p_{TiO_2} \approx 38.0$ mJ/m$^2$ vs. $\gamma^p_{SrTiO_3} \approx 35.7$ mJ/m$^2$). Thus, the explanation might involve the effect of CVD growth environments at 750~°C on the surface of rutile TiO$_2$, which is known to be sensitive to such high temperatures \cite{Wu.2020,Rogala.2019}. We hypothesize that defect sites might act as sparse but highly favorable nucleation centers, limiting the total nucleation density compared to the more structurally uniform binding sites on SrTiO$_3$. Once nucleated, the isolated domains can access a larger local supply of reactive species than densely nucleated regions, favoring lateral growth. Without competition from neighboring domains, these isolated islands can expand laterally to exceed 500~$\mu$m. This substrate--precursor interaction could explain the resulting star-like geometry, which is a classic hallmark of diffusion-limited anisotropic growth. Because the sparse domains grow rapidly, the expanding MoS$_2$ lattice encounters higher local concentration gradients than the flat edges, driving preferential corner growth and resulting in the observed star-like morphology \cite{Chen.2020}.

A third regime is observed on MgO, where nucleation is relatively sparse but results in large, dendritic MoS$_2$ domains extending over hundreds of micrometers (Figure~\ref{Fig:1}c iii), which eventually coalesce into a continuous layer (Figure S1c). The extracted monolayer and multilayer coverages are approximately 49\% and 28\%, respectively. As MgO is known to hydrate upon contact with aqueous solutions \cite{Brown.1999}, the precursor deposition and subsequent calcination may chemically modify the substrate surface. Such surface modification could then locally retain the liquid precursor and generate dense MoO$_3$-rich regions, promoting secondary nucleation and multilayer formation \cite{Park.2019}. When coupled with spatial variations in sulfur access across a modified surface, uniform two-dimensional growth and expansion may be hindered, yielding the observed dendritic architectures.

Finally, MoS$_2$ growth on SiC illustrates a low-coverage regime with many well-defined triangular domains and some large $>100~\mu$m domains. Driven by its comparatively low polar surface energy component ($\gamma^p = 10.90$~mJ/m$^2$), the relatively hydrophobic SiC surface ($\theta \approx 66.3^\circ$) is expected to promote partial precursor dewetting, which could lead to a low MoO$_3$ concentration after calcination. Consequently, the overall MoS$_2$ coverage remains lower than on the oxide substrates, with monolayer and multilayer coverages of approximately 15.1\% and 7.2\%, respectively (Figure S1d and S1f). Where localized precursor regions remain attached, the crystallographic structure of SiC may influence nucleation and domain orientation. Thus, the resulting domains exhibit the well-defined triangular shapes typically associated with more ordered MoS$_2$ growth conditions.

These observations indicate that, while nucleation in liquid-precursor CVD occurs across all investigated crystalline substrates, both the nucleation density and resulting growth morphology vary significantly. The substrate-dependent differences in wetting, high-temperature crystalline reconstruction, and chemical stability correlate with distinct MoS$_2$ domain sizes, densities, shapes, and multilayer fractions. The nanoscale origins of these pathways are examined in the following section.

\subsection*{Nanoscale Morphology and Phenomenological Growth Pathways}

\begin{figure*}[t]
\includegraphics[width=1\textwidth]{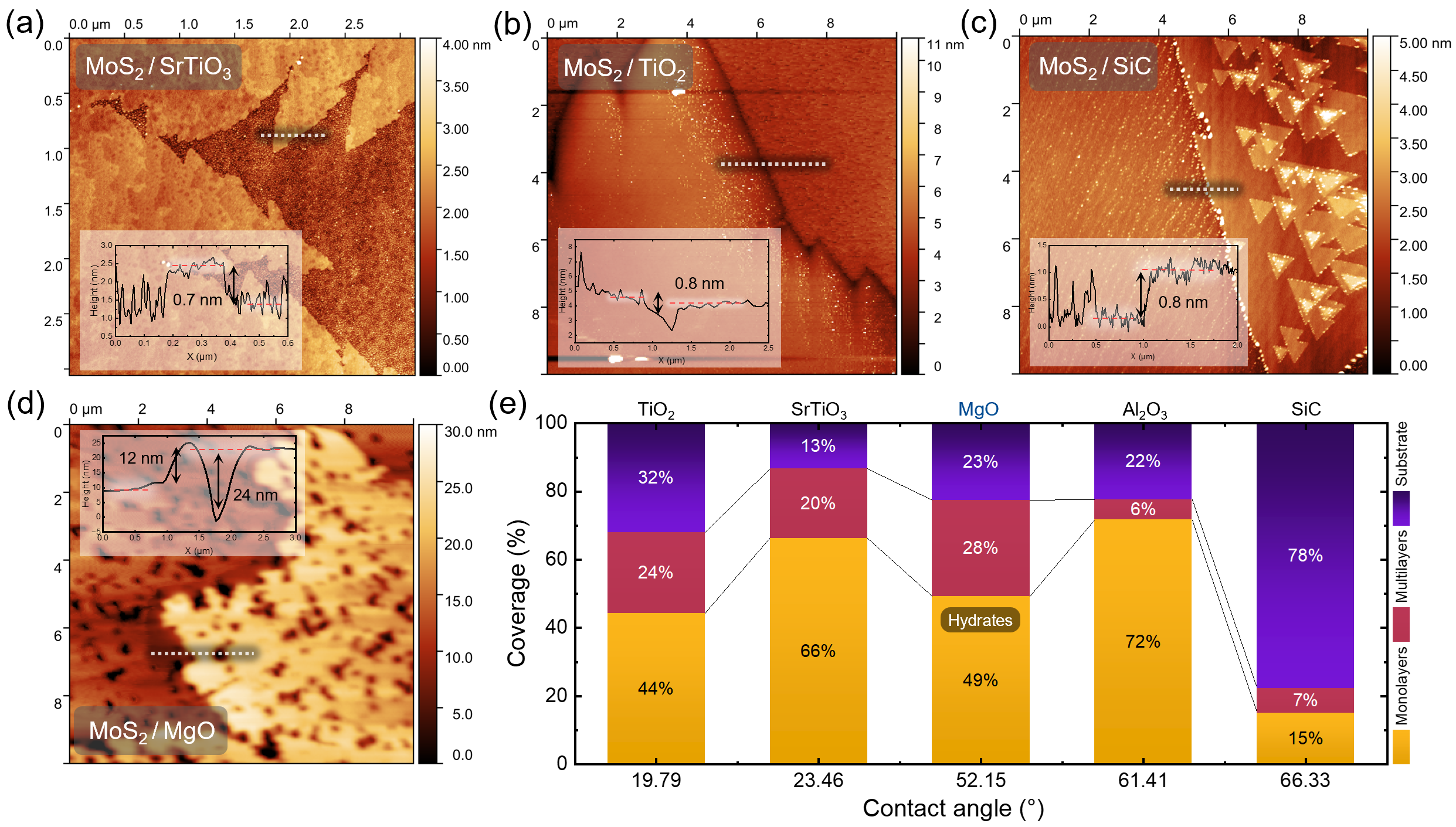}
\caption{AFM scans showing the CVD-grown MoS$_2$ layers on
\textbf{(a)} SrTiO$_3$,
\textbf{(b)} TiO$_2$,
\textbf{(c)} SiC, and
\textbf{(d)} MgO. A line scan in each AFM measurement indicates the MoS$_2$ height profile. 
\textbf{(e)} MoS$_2$ monolayer, multilayer, and uncovered-substrate fractions plotted against the measured substrate--water contact angle. The trend is interpreted as an empirical correlation between room-temperature wetting properties and the initial precursor distribution, with MgO representing a chemically distinct case due to likely surface modification during processing.
}
\label{Fig:2}
\end{figure*}

The AFM scans presented in Figure~\ref{Fig:2} provide insight into the substrate-dependent growth pathways. By characterizing the topography of the MoS$_2$ flakes and the substrate state after CVD, we correlate the observed morphological features with substrate-dependent precursor distribution, local conversion conditions, and effective mass transport. We note that the following interpretation is phenomenological: the present ex situ measurements do not directly determine sulfur transport, activation energies, or precursor binding energies.

As shown in Figure~\ref{Fig:2}a for MoS$_2$ on SrTiO$_3$, the MoS$_2$ film ($\approx$0.7~nm) exhibits step-like sub-layers ($\approx$0.2~nm) consisting of domains that coalesce into a leaf-like film at the edges. Similar sub-layer features and their relation to precursor conversion pathways have been reported in the literature \cite{Zhu.2017}. Since these steps are thinner than a full MoS$_2$ bilayer, their presence suggests incomplete or non-uniform sulfurization during growth. The conversion of MoO$_3$ to MoS$_2$ is known to proceed through intermediate MoO$_{3-x}$ species~\cite{Pondick.2018,Zhu.2017}, which could leave residual sub-oxide species or incompletely converted regions at grain boundaries under locally limited sulfur supply. This nanoscale morphology is consistent with the high-density, kinetically constrained growth observed optically.

In contrast, the growth of MoS$_2$ on TiO$_2$ reflects a highly anisotropic, non-equilibrium pathway. The AFM image (Figure~\ref{Fig:2}b) reveals well-defined MoS$_2$ flakes with straight edges terminating in star-like apices ($\approx$0.8~nm). This nanoscale topography is consistent with kinetically hindered edge attachment, where strong substrate--precursor interactions or local variations in the Mo:S ratio may affect growth into compact triangular domains \cite{Chen.2020}. Furthermore, the nanoscale protrusions observed on the flake surface indicate localized regions of incomplete conversion or residual precursor species, supporting the interpretation that the TiO$_2$ growth regime is strongly influenced by local precursor--substrate interaction.

Furthermore, the nucleation behavior on SiC translates into a growth pathway that appears influenced by the crystallographic and topographic structure of the substrate (Figure~\ref{Fig:2}c). The MoS$_2$ flakes ($\approx$0.8~nm) exhibit well-defined triangular geometries and an apparent orientational preference, consistent with possible step-edge-directed or crystallographically influenced growth. On 6H-SiC~(0001), characteristic step-bunching into half- or full-unit-cell terraces provides preferential nucleation sites and diffusion boundaries for the precursor \cite{Borovikov.2009,Kimoto.1997}, while the relatively smooth terraces between these steps may support extended lateral growth. This combination of localized nucleation and domain expansion may favor the formation of triangular domains characteristic of more near-equilibrium MoS$_2$ growth conditions \cite{Chen.2020}. 

MoS$_2$ grown on MgO exhibits severe morphological disruption with deep pits ($\approx$24~nm) extending through both the MoS$_2$ layer and the underlying substrate, as seen in Figure~\ref{Fig:2}d. The large height variation of the MoS$_2$ flake, around 12~nm relative to the surrounding MgO, indicates that the substrate surface is modified during processing. Previous studies indicate that aqueous precursors can locally hydrate MgO to form Mg(OH)$_2$~\cite{Yang.2025}, which subsequently decomposes during calcination~\cite{Kondo.2021}, potentially contributing to the observed craters. This heterogeneous, cratered surface provides a physical basis for the hindered two-dimensional expansion observed optically. MoS$_2$ growth occurring across this disrupted topography may lead to locally Mo-rich regions and the emergence of dendritic morphologies (Figure~\ref{Fig:1}c~iii).
To quantify the substrate morphological modification, pre-growth AFM characterization was performed on an as-received MgO substrate (Figure S2a,b). The analysis reveals that the pristine MgO surface already exhibits a rugged morphology, with an arithmetic mean roughness of Sa = 1.7~nm and a root-mean-square roughness of Sq = 2.1~nm (Figure S2c). After CVD growth, the roughness of the MoS$_2$-covered region increased to Sa = 2.6~nm and Sq = 3.5~nm, while the maximum pit depth (Sv) increased from 6.9~nm (pre-growth) to 17.7~nm (post-growth). This substantial increase in surface roughness is consistent with the hydration–dehydration pathway proposed above, where the aqueous precursor locally converts MgO to Mg(OH)$_2$, which subsequently decomposes during calcination, generating the observed craters.

\subsection*{Growth Framework}

To place these observations within a common framework, we describe the liquid-precursor CVD process phenomenologically as a substrate-influenced reaction--diffusion-like growth process. During sulfurization, the effective surface transport of sulfur-containing species can be represented by a characteristic diffusion length, $L_{\mathrm{S}} = \sqrt{D_{\mathrm{eff}}\tau}$, where $D_{\mathrm{eff}}$ is an effective substrate-dependent transport coefficient and $\tau$ is an effective residence time  \cite{Shendokar.2024}. Upon reaching Mo-rich reaction sites, conversion proceeds through thermally activated processes that depend on local surface chemistry, edge-attachment kinetics, and substrate topography. In a liquid-precursor system, the initial spatial distribution of the MoO$_3$ precursor is expected to be influenced by the wetting properties of the substrate, represented here by the contact angle $\theta$ and the polar/dispersive surface-energy components. Consequently, variations in substrate physics and chemistry may influence both the precursor distribution and the subsequent sulfurization pathway. We therefore conceptually separate the morphological evolution into an initial nucleation stage, influenced by precursor wetting and possible anchoring, and a subsequent lateral-growth stage, influenced by effective sulfur transport and local conversion kinetics.

Within this framework, Figure~\ref{Fig:2}e visualizes the empirical relationship between substrate wettability and final MoS$_2$ coverage. The highly polar and strongly wetting oxide surfaces, TiO$_2$ and SrTiO$_3$ (Table~\ref{tab:substrate_properties}), exhibit low contact angles and high polar surface-energy components. These conditions are expected to favor adhesion and retention of the aqueous precursor, resulting in relatively high MoS$_2$ coverage compared with the weakly wetting SiC surface. On SrTiO$_3$, this manifests as dense nucleation and limited lateral domain growth, consistent with a high density of distributed Mo-rich sites and a reduced effective lateral growth length. On TiO$_2$, the similarly strong wetting behavior is accompanied by sparse nucleation and large star-like domains, indicating that precursor spreading alone does not determine the final morphology. Instead, the TiO$_2$ growth pathway likely reflects an additional contribution from strong precursor--substrate interaction and anisotropic edge attachment.

In contrast, SiC has the largest contact angle and the lowest polar surface-energy component in the substrate library. This weaker wetting tendency is consistent with partial precursor dewetting and the much lower MoS$_2$ coverage observed on SiC. Where precursor material remains locally attached, topographic or crystallographic sites may still promote nucleation, leading to the well-defined triangular domains observed by AFM. MgO deviates from a simple wetting-controlled trend. Although its contact angle and polar surface-energy component are intermediate, its hydration-prone surface likely undergoes chemical and topographic restructuring during precursor deposition and calcination. This creates a chemically heterogeneous growth surface, promotes local precursor accumulation, and leads to elevated multilayer nucleation and dendritic growth.
Thus, Figure~\ref{Fig:2}e should be interpreted as an empirical correlation rather than a direct wetting law. The contact angle and surface-energy components provide useful descriptors of the initial precursor wetting tendency, whereas the final MoS$_2$ morphology also depends on substrate chemistry, precursor conversion, sulfur transport, and high-temperature surface restructuring.

These results highlight that morphology in liquid-phase precursor CVD is not governed by a universal growth mode, but instead reflects substrate-controlled variations in precursor distribution, local reaction conditions, and effective mass transport. These structural differences form the basis for the distinct optical responses discussed in the following section.

\subsection*{Spectroscopic Consequences}

\begin{figure*}[t]
\includegraphics[width=1\textwidth]{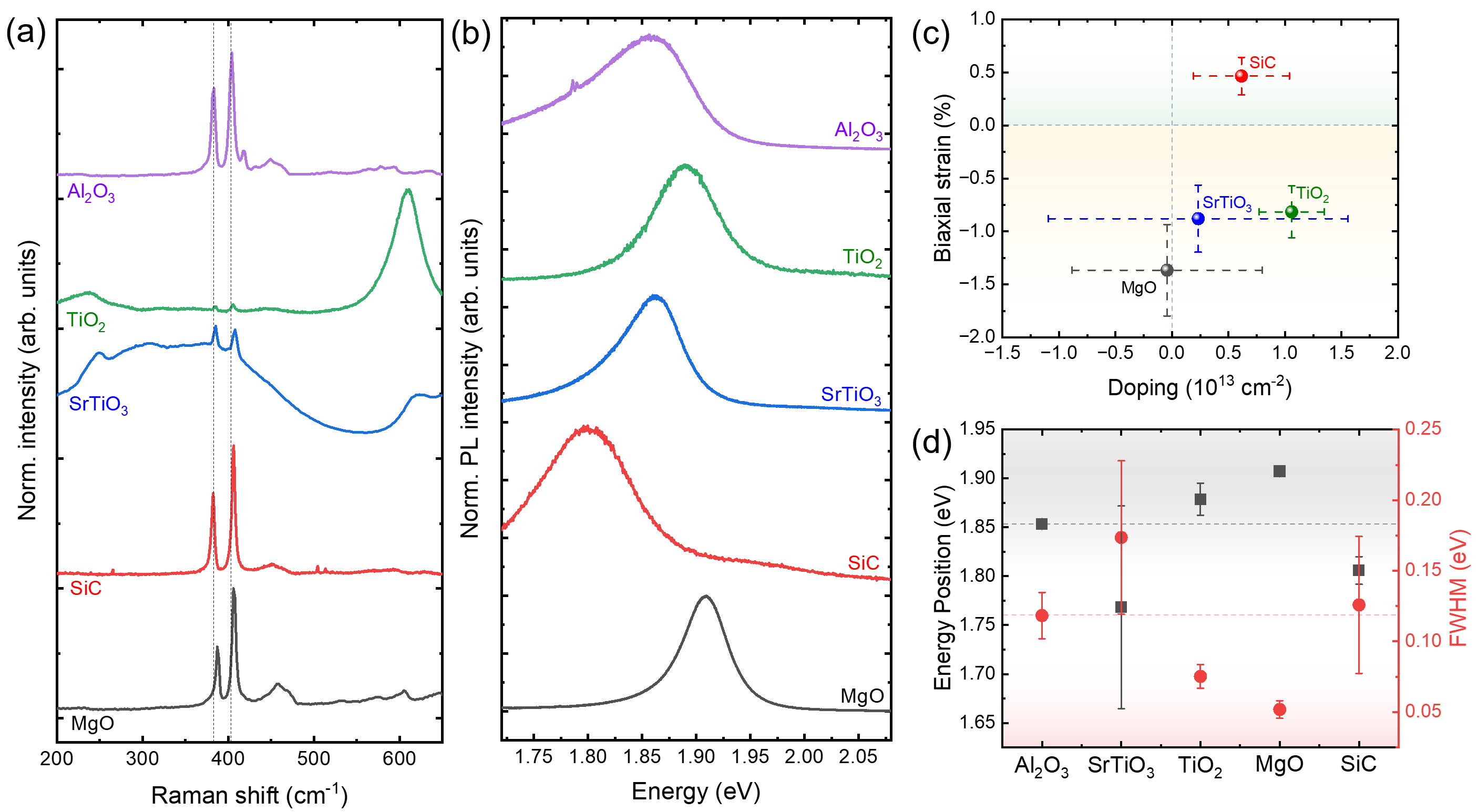}
\caption{
\textbf{(a)} Stacked representative Raman spectra of MoS$_2$ on sapphire, TiO$_2$, SrTiO$_3$, MgO, and SiC.
\textbf{(b)} Stacked representative PL spectra of MoS$_2$ on sapphire, TiO$_2$, SrTiO$_3$, MgO, and SiC.
\textbf{(c)} Raman-based relative indicators of biaxial strain and carrier-density shift with respect to sapphire-grown MoS$_2$.
\textbf{(d)} PL analysis comparing the peak energy positions and FWHM of the emission spectra, with sapphire-grown MoS$_2$ taken as the reference baseline.
}
\label{Fig:3}
\end{figure*}

The distinct substrate-defined growth pathways identified above correlate with modifications of the vibrational and optical properties of MoS$_2$. To investigate these effects, we employ Raman and PL spectroscopy as comparative probes of lattice strain, charge environment, and excitonic response. Raman and PL measurements were performed on visibly monolayer regions.

Raman spectroscopy provides insight into the coupling between the MoS$_2$ lattice and its local environment. In particular, the in-plane $E_{2g}^1$ mode is primarily sensitive to lattice strain~\cite{Conley.2013}, while the out-of-plane $A_{1g}$ mode is influenced by charge carrier density and interlayer interactions~\cite{Pollmann.2018,Daniel.2026}. Relative shifts of these modes with respect to a reference sample, here MoS$_2$ grown on sapphire, therefore provide a qualitative probe of substrate-induced perturbations. However, strain, doping, dielectric screening variations, and defect density can all contribute to the measured spectra~\cite{Pollmann.2018}. The extracted values should therefore be interpreted comparatively rather than as absolute strain and charge densities.

Figure~\ref{Fig:3}a shows the normalized Raman spectra across the substrate library. The mean fitted $E_{2g}^1$ and $A_{1g}$ modes are observed for MoS$_2$ on sapphire ($383.6\pm0.9~\text{cm}^{-1}$ and $406.6\pm2.7~\text{cm}^{-1}$), SrTiO$_3$ ($385.4\pm0.3~\text{cm}^{-1}$ and $406.4\pm2.8~\text{cm}^{-1}$), SiC ($382.4\pm0.3~\text{cm}^{-1}$ and $405.1\pm0.9~\text{cm}^{-1}$), and MgO ($386.5\pm0.7~\text{cm}^{-1}$ and $407.2\pm1.7~\text{cm}^{-1}$). On TiO$_2$, only weak MoS$_2$ Raman peaks can be resolved ($385\pm0.5~\text{cm}^{-1}$ and $404.5\pm0.7~\text{cm}^{-1}$), because the spectrum is dominated by the strong scattering response of the underlying substrate. The quantitative Raman analysis for TiO$_2$ should therefore be examined more critically. For each substrate, the reported values were obtained from at least four fitted Raman
spectra acquired from 4 flakes.

To estimate the relative strain and doping contributions, we utilize a vector-matrix analysis established by Pollmann \textit{et al.}~\cite{Pollmann.2020b}. The reference matrix, $T$, is given as

$$
T = 
\begin{pmatrix} 
-0.490\,\%/\text{cm}^{-1} & 0.073\,\%/\text{cm}^{-1} \\ 8.8 \times 10^{11}\,\text{cm}^{-2}/\text{cm}^{-1} & -4.64 \times  10^{12}\,\text{cm}^{-2}/\text{cm}^{-1} 
\end{pmatrix}.
$$
From this, the relative biaxial strain ($\Delta \varepsilon$) and carrier-density
shift ($\Delta n$) are estimated as
$$
\begin{pmatrix} \Delta \varepsilon \\ \Delta n \end{pmatrix} = T  \begin{pmatrix} \Delta\omega_E \\ \Delta\omega_A \end{pmatrix},
$$
where $\Delta\omega_E$ and $\Delta\omega_A$ are the frequency shifts relative to the sapphire baseline ($383.6\pm0.9~\text{cm}^{-1}$ and $406.6\pm2.7~\text{cm}^{-1}$). The transformation was applied to each fitted spectrum, after which the resulting values were averaged for each substrate, and the reported uncertainties denote the standard deviation across the individually fitted spectra. This approach provides an approximate separation of strain and doping contributions and assumes similar calibration across different substrates. In this convention, negative $\Delta n$ values represent relative p-type doping, while positive values indicate relative n-type doping with respect to the sapphire reference baseline. Substrate-dependent dielectric screening, defect concentration, and strain inhomogeneity may influence the absolute calibration of the Raman shifts. Thus, the extracted values in Figure~\ref{Fig:3}c are used as relative indicators of substrate-induced perturbation. 

For MoS$_2$ on SrTiO$_3$ ($a=3.90$~\AA), the Raman analysis indicates compressive strain of $-0.88\pm0.32$\%. The relative carrier-density shift, $(+0.23\pm1.33)\times10^{13}$~cm$^{-2}$, is unresolved within the uncertainty, so we make no reliable p- or n-type assignment.
Correspondingly, the PL spectrum is pronouncedly broadened, with a FWHM of $0.174\pm0.054$~eV and a red-shifted emission energy of $1.768\pm0.104$~eV compared to the reference MoS$_2$ grown on Al$_2$O$_3$ (PL emission at $1.853\pm0.005$~eV, FWHM of $0.118\pm0.016$~eV) (Figure~\ref{Fig:3}b,d). The red shift cannot be explained by the compressive-strain estimate alone and likely reflects competing contributions from morphological and structural inhomogeneity, local charge variations, defects, and the dielectric environment of SrTiO$_3$. The broad distribution of PL energies further indicates that several effects contribute simultaneously.

In contrast, only a weak MoS$_2$ Raman signature can be resolved on TiO$_2$ ($a=4.59$~\AA, $\epsilon_r\sim86$). The Raman analysis indicates compressive strain of $-0.81\pm0.25$\% and the clearest relative n-type carrier-density shift in the substrate library, $(+1.06\pm0.29)\times10^{13}$~cm$^{-2}$. However, these values should be interpreted cautiously because of the strong TiO$_2$ Raman background. The PL emission is relatively sharp and blue-shifted compared with sapphire, with a peak energy of $1.88\pm0.02$~eV and a FWHM of $0.08\pm0.008$~eV. The estimated compressive strain is consistent with a contribution to the higher emission energy, while the relative n-type shift, dielectric environment, and local defect landscape may also affect the excitonic response. Although MoS$_2$/TiO$_2$ interfaces are known to support interfacial charge transfer~\cite{Wei.2019}, the present steady-state PL and Raman data alone do not isolate this contribution.

A markedly different optical profile emerges for MoS$_2$ grown on MgO ($a=4.21$~\AA). Although the dielectric constant of MgO ($\epsilon_r\sim9.8$) is close to that of the sapphire reference, both Raman and PL results differ from the sapphire baseline. The Raman analysis indicates pronounced compressive strain of $-1.36\pm0.43$\%, while the relative carrier-density shift of $(-0.04\pm0.84)\times10^{13}$~cm$^{-2}$ is indistinguishable from zero within the uncertainty. The Raman analysis therefore does not support a resolved p- or n-type carrier-density shift for MoS$_2$ on MgO. Despite the highly disrupted, cratered morphology observed by AFM, the PL emission is sharper and significantly blue-shifted. The strong compressive strain is consistent with a contribution to this blue shift, while the chemically modified MgO surface, structural heterogeneity, and local defect variations may also influence the optical response.

Finally, MoS$_2$ grown on SiC ($a=3.08$~\AA, $\epsilon_r\sim9.7$) exhibits tensile strain of $+0.47\pm0.18$\% according to the Raman analysis. The relative carrier-density shift is $(+0.62\pm0.42)\times10^{13}$~cm$^{-2}$, indicating a relative n-type doping trend. The PL spectrum is red-shifted ($1.806\pm0.014$~eV, FWHM $0.126\pm0.048$~eV), and the tensile-strain estimate is consistent with a contribution to the lower emission energy. However, local defects, morphology-dependent interactions, charge redistribution, and dielectric effects may also influence the optical response. The extracted values are therefore used only as comparative indicators and should not be interpreted as absolute strain or carrier density.

Overall, the Raman analysis reveals distinct substrate-dependent perturbations rather than a uniform strain or carrier-density trend. MoS$_2$ on SiC exhibits a tensile-strain estimate, whereas SrTiO$_3$, TiO$_2$, and MgO exhibit compressive strain. TiO$_2$ shows the clearest relative n-type carrier-density shift, SiC shows a weaker positive trend, and the corresponding shifts for SrTiO$_3$ and MgO are unresolved within their uncertainties. These trends do not provide a single explanation for the PL response. The strain estimates are consistent with contributions to the red-shifted emission on SiC and the blue-shifted emission on TiO$_2$ and MgO, whereas the strongly red-shifted SrTiO$_3$ emission cannot be explained by its compressive-strain estimate alone. The optical response therefore likely reflects combined contributions from strain, charge environment, morphological and structural inhomogeneity, defects, and dielectric screening. This reinforces the conclusion that the substrate acts as an active parameter influencing both the growth morphology and optical properties of MoS$_2$.

\section{Conclusion}

In summary, we demonstrate that the scalable integration of MoS$_2$ onto functional substrates via liquid-phase precursor CVD is strongly influenced by the substrate’s chemical and physical properties. This behavior is consistent with a growth process in which the spatial distribution of the pre-deposited Mo-containing precursor and substrate-dependent surface processes contribute strongly to the subsequent sulfurization and lateral growth.

Thus, the observed growth behavior appears to depend on the interplay between interfacial surface wetting and local sulfur mass transport. Using a diverse substrate library, we identify distinct substrate-dependent growth regimes consistent with reduced effective lateral growth on SrTiO$_3$, possible precursor anchoring and anisotropic edge attachment on TiO$_2$, hydration-related surface restructuring on MgO, and possible step-edge or crystallographic influence on SiC. By modifying these local reaction--diffusion environments, the substrates promote distinct morphological adaptations that correlate with measurable variations in lattice strain and, for some substrates, relative carrier-density indicators, together with differences in the dielectric environment. Collectively, these substrate-induced perturbations likely contribute to the observed excitonic response. While substrate-dependent strain and charge-environment changes may contribute to some of the observed PL shifts, they do not consistently explain all substrates, particularly SrTiO$_3$. Variations in grain-boundary density, local defect concentration, morphology, and dielectric screening may also influence the optical response.

While the proposed reaction-diffusion pathways are phenomenological and based on ex situ characterization, they provide a useful framework for tuning the structural and optical properties of large-area 2D materials. Given their potential to reduce sensitivity to the spatial delivery of the Mo precursor, CVD methods utilizing partially pre-deposited precursors offer a potentially scalable pathway for 2D TMDC synthesis on functional substrates \cite{Yang.2017,Robertson.2019}, underscoring the critical need to further investigate these constrained solid-state growth dynamics.

\section{Methods}
\label{methods}

\subsection*{Sample fabrication}

The crystalline Sapphire~(0001), TiO$_2$~(001), SrTiO$_3$~(100), MgO~(100), and SiC~(0001) substrates utilized in this work were commercially acquired from \textit{Crystec GmbH} with dimensions of 1~cm $\times$ 1~cm $\times$ 1~mm.

To synthesize MoS$_2$, a liquid-phase molybdenum-containing precursor (an aqueous solution comprising 1~mL of 100\% saturated ammonium heptamolybdate (AHM, \textit{Sigma Aldrich}), 1~mL of OptiPrep (\textit{Sigma Aldrich}), and 3.5~mL of deionized water) was spin-coated onto the substrates~(12,000~rpm, 45~s). Following deposition, a calcination step was performed by annealing the substrates at 300~$^\circ$C for 30~minutes under ambient conditions to thermally decompose the AHM and form solid molybdenum trioxide (MoO$_3$) intermediates across the substrate surfaces. 2\% sodium cholate aqueous solution (\textit{Sigma Aldrich}) was applied via spin-coating (4500~rpm, 45~s) to act as a seeding promoter.

The subsequent sulfurization process was conducted in a three-zone tube furnace (\textit{ThermConcept}, ROK 70/750/12-3z). As illustrated in Figure~\ref{Fig:1}b, 100~mg of sulfur powder was placed in a combustion boat situated in the upstream zone (Zone 1), which was heated to 170~°C. A regulated argon (Ar) carrier gas flow of 500~sccm was established through the quartz tube to transport the vaporized sulfur downstream into the central reaction zone (Zone 2). Here, the pre-calcined substrates were heated to 750~°C for 30~minutes to drive the solid-state conversion of the MoO$_3$ into crystalline molybdenum disulfide (MoS$_2$). The downstream exhaust zone (Zone 3) was maintained at 250~°C to prevent the immediate condensation of reaction by-products and ensure stable gas dynamics within the reactor.

\subsection*{Optical Measurements}

Confocal PL and Raman measurements were conducted using a \textit{Renishaw} “inVia Raman microscope” with a laser wavelength of 532~nm and a grating with 1800~g/mm. At least four spectra were acquired with a high signal-to-noise ratio for PL and Raman, respectively. Additional PL and Raman measurements were conducted using a \textit{WITec alpha300 RA} confocal Raman spectrometer, along with a laser wavelength of 532~nm and a grating with 1800~g/mm. A laser power of roughly 1.25~$\times$~10$^5$ W/cm$^2$ was used.
The software \textit{OriginPro} was used to fit and analyze the Raman and PL data. 
Coverage percentages were determined via threshold-based image segmentation of ten 200~$\mu$m x 200~$\mu$m optical micrographs per substrate using \textit{ImageJ}.

\subsection*{Atomic Force Microscopy}

AFM measurements were performed using a \textit{Park Systems NX10} in non-contact mode.
Topography was acquired in the attractive regime via closed-loop amplitude control near the cantilever resonance. A conductive \textit{Park Systems AC160TS} probe (resonance frequency 260$–$300~kHz, quality factor $Q = 201–530$, tip radius $\sim$7~nm) was used, enabling sensitive detection of long-range interactions. Images were acquired at scan rates of $0.15–0.5$~Hz with a lateral resolution of $256–1024$~pixels~per~line.
To characterize the bare reference MgO substrate before growth, topography was measured using the integrated AFM module of a \textit{WITec alpha300 RA} confocal spectrometer. Measurements were performed in AC (tapping) mode at a cantilever drive frequency of 83.65~kHz (tip radius $\sim$10~nm). Topographic maps ($1 \times 1~\mu$m$^2$) were acquired at a lateral resolution of $100 \times 100$~pixels with a trace/retrace time of 4.0~s per line (scan rate of 0.125~Hz). The AFM data analysis and visualization were done using \textit{Gwyddion~2.66}. 

\subsection*{Surface Energy Measurements}
The macroscopic wettability and surface free energy (SFE) of the pristine substrates were characterized using a \textit{KRÜSS Mobile Surface Analyzer}. 
Measurements were performed using the double-sessile drop technique. For each substrate, precisely dosed droplets ($\approx 1~\mu$L) of two distinct probe liquids—deionized water (highly polar) and diiodomethane (highly dispersive)—were deposited and photographed to determine their static contact angles. The optical data were analyzed using the \textit{ADVANCE\textsuperscript{\tiny\textregistered}} software. Subsequently, the extracted contact angles were fitted using the Owens-Wendt-Rabel-Kaelble (OWRK) method to calculate the total surface free energy ($\gamma_{tot}$) and decouple it into its fundamental dispersive ($\gamma^d$) and polar ($\gamma^p$) components.


\section*{Data Availability}
The data supporting the findings of this work are publicly available under:

\url{https://doi.org/10.17172/nomad.cjxr-h6yv}.

\section*{Conflicts of interest}
There are no conflicts to declare.

\section*{Acknowledgments}
O.K., C.S., and M.S. acknowledge financial support from the DFG within the IRTG 2803: 2D Mature, project No. 461605777, as well as under project No. 429784087. We thank the BMBF for the financial support under Project No. 05K16PG1. O.K. acknowledges support from Hanna Mueller with AFM measurements. This work was partially supported by the research platform PULVECHAMP funded by the Region Normandie.

\bibliography{apssamp}

@article{Rogala.2019,
 author = {Rogala, M. and Bihlmayer, G. and Dabrowski, P. and Rodenb{\"u}cher, C. and Wrana, D. and Krok, F. and Klusek, Z. and Szot, K.},
 year = {2019},
 title = {Self-reduction of the native TiO2 (110) surface during cooling after thermal annealing - in-operando investigations},
 pages = {12563},
 volume = {9},
 number = {1},
 journal = {Scientific reports},
 doi = {\url{10.1038/s41598-019-48837-3}}
}

@article{Wu.2020,
 author = {Wu, Longxia and Wang, Zhengming and Xiong, Feng and Sun, Guanghui and Chai, Peng and Zhang, Zhen and Xu, Hong and Fu, Cong and Huang, Weixin},
 year = {2020},
 title = {Surface chemistry and photochemistry of small molecules on rutile TiO2(001) and TiO2(011)-(2 $\times$ 1) surfaces: The crucial roles of defects},
 pages = {044702},
 volume = {152},
 number = {4},
 journal = {The Journal of chemical physics},
 doi = {\url{10.1063/1.5135945}}
}

@article{Borovikov.2009,
 author = {Borovikov, Valery and Zangwill, Andrew},
 year = {2009},
 title = {Step bunching of vicinal 6H-SiC{0001} surfaces},
 volume = {79},
 number = {24},
 issn = {1098-0121},
 journal = {Physical Review B},
 doi = {\url{10.1103/PhysRevB.79.245413}}
}

@article{Kimoto.1997,
 author = {Kimoto, T. and Itoh, A. and Matsunami, H.},
 year = {1997},
 title = {Step-Controlled Epitaxial Growth of High-Quality SiC Layers},
 pages = {247--262},
 volume = {202},
 number = {1},
 issn = {0370-1972},
 journal = {physica status solidi (b)},
 doi = {\url{10.1002/1521-3951(199707)202:1{\%}3C247::AID-PSSB247{\%}3E3.0.CO;2-Q}}
}

@article{Perry.1997,
 author = {Perry, Scott S. and Merrill, Philip B.},
 year = {1997},
 title = {Preparation and characterization of MgO(100) surfaces},
 pages = {268--276},
 volume = {383},
 number = {2-3},
 issn = {00396028},
 journal = {Surface Science},
 doi = {\url{10.1016/S0039-6028(97)00185-4}}
}

@article{Soares.2002,
 author = {Soares, E. A. and {van Hove}, M. A. and Walters, C. F. and McCarty, K. F.},
 year = {2002},
 title = {Structure of the \textgreek{a}$-$Al2O3(0001) surface from low-energy electron diffraction: Al termination and evidence for anomalously large thermal vibrations},
 volume = {65},
 number = {19},
 issn = {1098-0121},
 journal = {Physical Review B},
 doi = {\url{10.1103/PhysRevB.65.195405}}
}

@article{Starke.1997,
 author = {Starke, U.},
 year = {1997},
 title = {Atomic Structure of Hexagonal SiC Surfaces},
 pages = {475--499},
 volume = {202},
 number = {1},
 issn = {0370-1972},
 journal = {physica status solidi (b)},
 doi = {\url{10.1002/1521-3951(199707)202:1{\%}3C475::AID-PSSB475{\%}3E3.0.CO;2-E}}
}

@article{Erdman.2002,
 author = {Erdman, Natasha and Poeppelmeier, Kenneth R. and Asta, Mark and Warschkow, Oliver and Ellis, Donald E. and Marks, Laurence D.},
 year = {2002},
 title = {The structure and chemistry of the TiO(2)-rich surface of SrTiO(3) (001)},
 pages = {55--58},
 volume = {419},
 number = {6902},
 issn = {0028-0836},
 journal = {Nature},
 doi = {\url{10.1038/nature01010}}
}

@article{Diebold.2003,
 author = {Diebold, Ulrike},
 year = {2003},
 title = {The surface science of titanium dioxide},
 pages = {53--229},
 volume = {48},
 number = {5-8},
 issn = {01675729},
 journal = {Surface Science Reports},
 doi = {\url{10.1016/S0167-5729(02)00100-0}}
}

@article{Kharsah.2026b,
 author = {Kharsah, Osamah and von Kuczkowski, Jonah and Shahin, Ahmed and Albalkhi, Rouaa and Al-Amwi, Aya and Titi, Ibrahim and Liebsch, Yossarian and Altenhoff, Oliver and Hagemann, Ulrich and Hierzenberger, Anke and Musselman, Kevin and Schleberger, Marika},
 year = {2026},
 title = {{Anomalous Sub-band Gap Emission at Room Temperature in Mixed-Dimensional {WS$_2$}/{Cu$_2$O} Heterostructures}},
 pages = {e75662},
 journal = {Small},
 doi = {10.1002/smll.75662}
}

@article{Daniel.2026,
 author = {Daniel, Leon and Liebsch, Yossarian and Lintz, Charleen and Javed, Umair and Kharsah, Osamah and Breuer, Lars and Kotakoski, Jani and Schleberger, Marika},
 year = {2026},
 title = {In situ Raman study on sulfur vacancies in monolayer MoS 2},
 pages = {025033},
 volume = {13},
 number = {2},
 journal = {2D Materials},
 doi = {\url{10.1088/2053-1583/ae6b2b}}
}

@article{Kharsah.2026,
 author = {Kharsah, Osamah and Sutarma, Dedi and Daniel, Leon and Vidish, Denys and Zheng, Ruofei and Netzke, Sam and Hagemann, Ulrich and Morgado, Felipe F. and Abdelbaky, Mohamed and Ustenko, Sofiia and Mertin, Wolfgang and Bacher, Gerd and Sciaini, Germ{\'a}n and Musselman, Kevin and Kratzer, Peter and Schleberger, Marika},
 year = {2026},
 title = {ZnO Surface Morphology and Chemistry Govern Exciton Dynamics in {WS$_2$}/ZnO Heterostructures},
 pages = {5077--5089},
 volume = {8},
 number = {12},
 issn = {2637-6113},
 journal = {ACS Applied Electronic Materials},
 doi = {\url{10.1021/acsaelm.6c00496}}
}

@book{INSPECInformationservice.1995,
 year = {1995},
 title = {Properties of silicon carbide},
 url = {\url{http://purl.oclc.org/DLF/benchrepro0212}},
 address = {London, U.K.},
 volume = {no. 13},
 publisher = {{IEE INSPEC}},
 isbn = {9780852968703},
 series = {EMIS datareviews series}
}

@article{Brown.1999,
 author = {Brown, Gordon E. and Henrich, Victor E. and Casey, William H. and Clark, David L. and Eggleston, Carrick and Felmy, Andrew and Goodman, D. Wayne and Gr{\"a}tzel, Michael and Maciel, Gary and McCarthy, Maureen I. and Nealson, Kenneth H. and Sverjensky, Dimitri A. and Toney, Michael F. and Zachara, John M.},
 year = {1999},
 title = {Metal Oxide Surfaces and Their Interactions with Aqueous Solutions and Microbial Organisms},
 pages = {77--174},
 volume = {99},
 number = {1},
 journal = {Chemical reviews},
 doi = {\url{10.1021/cr980011z}}
}

@article{Fontanella.1974,
 author = {Fontanella, John and Andeen, Carl and Schuele, Donald},
 year = {1974},
 title = {Low-frequency dielectric constants of \textgreek{a}-quartz, sapphire, MgF2, and MgO},
 pages = {2852--2854},
 volume = {45},
 number = {7},
 issn = {0021-8979},
 journal = {Journal of Applied Physics},
 doi = {\url{10.1063/1.1663690}}
}

@article{Hazen.1976,
 author = {Hazen, R. M.},
 year = {1976},
 title = {Effects of temperature and pressure on the cell dimension and X-ray temperature factors of periclase},
 pages = {266--271},
 volume = {61},
 number = {3-4},
 issn = {0003-004X},
 journal = {American Mineralogist}
}

@article{Parker.1961,
 author = {Parker, Rebecca A.},
 year = {1961},
 title = {Static Dielectric Constant of Rutile (Ti O2 ), 1.6-1060°K},
 pages = {1719--1722},
 volume = {124},
 number = {6},
 issn = {0031-899X},
 journal = {Physical Review},
 doi = {\url{10.1103/PhysRev.124.1719}}
}

@article{Cromer.1955,
 author = {Cromer, Don T. and Herrington, K.},
 year = {1955},
 title = {The Structures of Anatase and Rutile},
 pages = {4708--4709},
 volume = {77},
 number = {18},
 issn = {0002-7863},
 journal = {Journal of the American Chemical Society},
 doi = {\url{10.1021/ja01623a004}}
}

@article{Kawasaki.2017,
 author = {Kawasaki, Seiji and Holmstr{\"o}m, Eero and Takahashi, Ryota and Spijker, Peter and Foster, Adam S. and Onishi, Hiroshi and Lippmaa, Mikk},
 year = {2017},
 title = {Intrinsic Superhydrophilicity of Titania-Terminated Surfaces},
 pages = {2268--2275},
 volume = {121},
 number = {4},
 issn = {1932-7447},
 journal = {The Journal of Physical Chemistry C},
 doi = {\url{10.1021/acs.jpcc.6b12130}}
}

@article{Tumarkin.2024,
 author = {Tumarkin, Andrey and Sapego, Eugene and Gagarin, Alexander and Bogdan, Alexey and Karamov, Artem and Serenkov, Igor and Sakharov, Vladimir},
 year = {2024},
 title = {SrTiO3 Thin Films on Dielectric Substrates for Microwave Applications},
 pages = {3},
 volume = {14},
 number = {1},
 journal = {Coatings},
 doi = {\url{10.3390/coatings14010003}}
}

@article{Solokha.2019,
 author = {Solokha, Vladyslav and Garai, Debi and Wilson, Axel and Duncan, David A. and Thakur, Pardeep K. and Hingerl, Kurt and Zegenhagen, J{\"o}rg},
 year = {2019},
 title = {Water Splitting on Ti-Oxide-Terminated SrTiO 3 (001)},
 pages = {17232--17238},
 volume = {123},
 number = {28},
 issn = {1932-7447},
 journal = {The Journal of Physical Chemistry C},
 doi = {\url{10.1021/acs.jpcc.9b01730}}
}

@article{Smink.2024,
 author = {Smink, Sander and Majer, Lena N. and Boschker, Hans and Mannhart, Jochen and Braun, Wolfgang},
 year = {2024},
 title = {Long-Range Atomic Order on Double-Stepped Al2O3(0001) Surfaces},
 pages = {e2312899},
 volume = {36},
 number = {24},
 journal = {Advanced materials (Deerfield Beach, Fla.)},
 doi = {\url{10.1002/adma.202312899}}
}

@article{Lee.1985,
 author = {Lee, W. E. and Lagerlof, K. P. D.},
 year = {1985},
 title = {Structural and electron diffraction data for sapphire (\textgreek{a}--al 2 o 3 )},
 pages = {247--258},
 volume = {2},
 number = {3},
 issn = {0741-0581},
 journal = {Journal of Electron Microscopy Technique},
 doi = {\url{10.1002/jemt.1060020309}}
}

@article{Robertson.2019,
 author = {Robertson, John and Blomdahl, Daniel and Islam, Kazi and Ismael, Timothy and Woody, Maxwell and Failla, Jacqueline and Johnson, Michael and Zhang, Xiaodong and Escarra, Matthew},
 year = {2019},
 title = {Rapid-throughput solution-based production of wafer-scale 2D MoS2},
 volume = {114},
 number = {16},
 issn = {0003-6951},
 journal = {Applied Physics Letters},
 doi = {\url{10.1063/1.5093039}}
}

@article{Yang.2017,
 author = {Yang, Heeseung and Giri, Anupam and Moon, Sungmin and Shin, Sangbae and Myoung, Jae-Min and Jeong, Unyong},
 year = {2017},
 title = {Highly Scalable Synthesis of MoS 2 Thin Films with Precise Thickness Control via Polymer-Assisted Deposition},
 pages = {5772--5776},
 volume = {29},
 number = {14},
 issn = {0897-4756},
 journal = {Chemistry of Materials},
 doi = {\url{10.1021/acs.chemmater.7b01605}}
}

@article{Zhu.2017,
 author = {Zhu, Dancheng and Shu, Haibo and Jiang, Feng and Lv, Danhui and Asokan, Vijayshankar and Omar, Omar and Yuan, Jun and Zhang, Ze and Jin, Chuanhong},
 year = {2017},
 title = {Capture the growth kinetics of CVD growth of two-dimensional MoS2},
 volume = {1},
 number = {1},
 journal = {npj 2D Materials and Applications},
 doi = {\url{10.1038/s41699-017-0010-x}}
}

@article{Pondick.2018,
 author = {Pondick, Joshua V. and Woods, John M. and Xing, Jie and Zhou, Yu and Cha, Judy J.},
 year = {2018},
 title = {Stepwise Sulfurization from MoO 3 to MoS 2 via Chemical Vapor Deposition},
 pages = {5655--5661},
 volume = {1},
 number = {10},
 issn = {2574-0970},
 journal = {ACS Applied Nano Materials},
 doi = {\url{10.1021/acsanm.8b01266}}
}

@article{Shendokar.2024,
 author = {Shendokar, Sachin and Hossen, Moha Feroz and Aravamudhan, Shyam},
 year = {2024},
 title = {Wafer-Scale ALD Synthesis of MoO3 Sulfurized to MoS2},
 pages = {673},
 volume = {14},
 number = {8},
 journal = {Crystals},
 doi = {\url{10.3390/cryst14080673}}
}

@article{Seravalli.2021,
 author = {Seravalli, Luca and Bosi, Matteo},
 year = {2021},
 title = {A Review on Chemical Vapour Deposition of Two-Dimensional MoS2 Flakes},
 volume = {14},
 number = {24},
 issn = {1996-1944},
 journal = {Materials (Basel, Switzerland)},
 doi = {\url{10.3390/ma14247590}}
}

@article{Tang.2020,
 author = {Tang, Lei and Li, Tao and Luo, Yuting and Feng, Simin and Cai, Zhengyang and Zhang, Hang and Liu, Bilu and Cheng, Hui-Ming},
 year = {2020},
 title = {Vertical Chemical Vapor Deposition Growth of Highly Uniform 2D Transition Metal Dichalcogenides},
 pages = {4646--4653},
 volume = {14},
 number = {4},
 journal = {ACS nano},
 doi = {\url{10.1021/acsnano.0c00296}}
}

@article{Conley.2013,
 author = {Conley, Hiram J. and Wang, Bin and Ziegler, Jed I. and Haglund, Richard F. and Pantelides, Sokrates T. and Bolotin, Kirill I.},
 year = {2013},
 title = {Bandgap engineering of strained monolayer and bilayer MoS2},
 pages = {3626--3630},
 volume = {13},
 number = {8},
 journal = {Nano letters},
 doi = {\url{10.1021/nl4014748}}
}

@article{Kondo.2021,
 author = {Kondo, Atsushi and Kurosawa, Ryo and Ryu, Junichi and Matsuoka, Masaya and Takeuchi, Masato},
 year = {2021},
 title = {Investigation on the Mechanisms of Mg(OH) 2 Dehydration and MgO Hydration by Near-Infrared Spectroscopy},
 pages = {10937--10947},
 volume = {125},
 number = {20},
 issn = {1932-7447},
 journal = {The Journal of Physical Chemistry C},
 doi = {\url{10.1021/acs.jpcc.1c01470}}
}

@article{Yang.2025,
 author = {Yang, Peng and Bracco, Jacquelyn N. and {Camacho Meneses}, Gabriela and Yuan, Ke and Stubbs, Joanne E. and Boamah, Mavis D. and Brahlek, Matthew and Sassi, Michel and Eng, Peter J. and Boebinger, Matthew G. and Borisevich, Albina and Wanhala, Anna K. and Wang, Zheming and Rosso, Kevin M. and Stack, Andrew G. and Weber, Juliane},
 year = {2025},
 title = {Carbonation of MgO Single Crystals: Implications for Direct Air Capture of CO2},
 pages = {3484--3494},
 volume = {59},
 number = {7},
 journal = {Environmental science {\&} technology},
 doi = {\url{10.1021/acs.est.4c09713}}
}

@article{Chen.2020,
 author = {Chen, Fei and Su, Weitao and Zhao, Shichao and Lv, Yanfei and Ding, Su and Fu, Li},
 year = {2020},
 title = {Morphological evolution of atomically thin MoS 2 flakes synthesized by a chemical vapor deposition strategy},
 pages = {4174--4179},
 volume = {22},
 number = {24},
 journal = {CrystEngComm},
 doi = {\url{10.1039/D0CE00558D}}
}

@article{Kalt.2023,
 author = {Kalt, Romana Alice and Arcifa, Andrea and W{\"a}ckerlin, Christian and Stemmer, Andreas},
 year = {2023},
 title = {CVD of MoS2 single layer flakes using Na2MoO4 - impact of oxygen and temperature-time-profile},
 pages = {18871--18882},
 volume = {15},
 number = {46},
 journal = {Nanoscale},
 doi = {\url{10.1039/D3NR03907B}}
}

@article{Kandybka.2024,
 author = {Kandybka, Iryna and Groven, Benjamin and {Medina Silva}, Henry and Sergeant, Stefanie and {Nalin Mehta}, Ankit and Koylan, Serkan and Shi, Yuanyuan and Banerjee, Sreetama and Morin, Pierre and Delabie, Annelies},
 year = {2024},
 title = {Chemical Vapor Deposition of a Single-Crystalline MoS2 Monolayer through Anisotropic 2D Crystal Growth on Stepped Sapphire Surface},
 pages = {3173--3186},
 volume = {18},
 number = {4},
 journal = {ACS nano},
 doi = {\url{10.1021/acsnano.3c09364}}
}

@article{GovindRajan.2016,
 author = {{Govind Rajan}, Ananth and Warner, Jamie H. and Blankschtein, Daniel and Strano, Michael S.},
 year = {2016},
 title = {Generalized Mechanistic Model for the Chemical Vapor Deposition of 2D Transition Metal Dichalcogenide Monolayers},
 pages = {4330--4344},
 volume = {10},
 number = {4},
 journal = {ACS nano},
 doi = {\url{10.1021/acsnano.5b07916}}
}

@article{Huang.2026,
 author = {Huang, Gang and Chen, Ruosi and Chen, Mingxi and Chen, Xianfeng and Jiang, Mengting and Xing, Yu and Wang, Jiang and Liang, Boqun and Liu, Qiushi and Li, Xiangdong and Lau, Chit Siong and Dong, Xiaonan and Agarwal, Piyush and Ke, Lin and Assad, Syed M. and Soh, Jian-Rui and Lourembam, James and Cho, Young-Wook and Liang, Qingcheng and Li, Jian and Zhang, Xiao and Ma, Yuan and Lu, Yuerui and Lam, Ping Koy and Ma, Xuezhi},
 year = {2026},
 title = {Transfer and beyond: emerging strategies and trends in two-dimensional material device fabrication},
 pages = {2574--2634},
 volume = {55},
 number = {5},
 journal = {Chemical Society reviews},
 doi = {\url{10.1039/D5CS00531K}}
}

@article{Shen.2021,
 author = {Shen, Ying--Chun and Wu, Yu--Ting and Lee, Ling and Chen, Jyun--Hong and Wani, Sumayah Shakil and Yang, Tzu--Yi and Luo, Chih Wei and Siao, Ming--Deng and Yu, Yi--Jen and Chiu, Po--Wen and Chueh, Yu--Lun},
 year = {2021},
 title = {Rational Design on Wrinkle--Less Transfer of Transition Metal Dichalcogenide Monolayer by Adjustable Wettability--Assisted Transfer Method},
 volume = {31},
 number = {45},
 issn = {1616-301X},
 journal = {Advanced Functional Materials},
 doi = {\url{10.1002/adfm.202104978}}
}

@article{An.2022,
 author = {An, Gwang Hwi and {Jin Kim}, Su and Kim, Sanghyeon and Shin, So Jeong and Choi, Min and Kim, Dohyun and Rahman, Ikhwan Nur and Bang, Junhyeok and Kim, Kyungwan and Kim, Dong-Hyun and Lee, Hyun Seok},
 year = {2022},
 title = {Growth mode control of CVD-grown WS2 monolayer flakes via O2 pre-annealing for organic surfactant oxidation},
 pages = {152564},
 volume = {585},
 issn = {01694332},
 journal = {Applied Surface Science},
 doi = {\url{10.1016/j.apsusc.2022.152564}}
}

@article{Esposito.2023,
 author = {Esposito, F. and Bosi, M. and Attolini, G. and Rossi, F. and Panasci, S. E. and Fiorenza, P. and Giannazzo, F. and Fabbri, F. and Seravalli, L.},
 year = {2023},
 title = {Role of density gradients in the growth dynamics of 2-dimensional MoS2 using liquid phase molybdenum precursor in chemical vapor deposition},
 pages = {158230},
 volume = {639},
 issn = {01694332},
 journal = {Applied Surface Science},
 doi = {\url{10.1016/j.apsusc.2023.158230}}
}

@article{Hou.2024,
 author = {Hou, Yuan and Zhou, Jingzhuo and He, Zezhou and Chen, Juzheng and Zhu, Mengya and Wu, HengAn and Lu, Yang},
 year = {2024},
 title = {Tuning instability in suspended monolayer 2D materials},
 pages = {4033},
 volume = {15},
 number = {1},
 journal = {Nature communications},
 doi = {\url{10.1038/s41467-024-48345-7}}
}

@article{Xi.2014,
 author = {Xi, Jinyang and Zhao, Tianqi and Wang, Dong and Shuai, Zhigang},
 year = {2014},
 title = {Tunable Electronic Properties of Two-Dimensional Transition Metal Dichalcogenide Alloys: A First-Principles Prediction},
 pages = {285--291},
 volume = {5},
 number = {2},
 journal = {The journal of physical chemistry letters},
 doi = {\url{10.1021/jz402375s}}
}

@article{Xiao.2020,
 author = {Xiao, Yifan and Min, Long and Liu, Xinke and Liu, Wenjun and Younis, Usman and Peng, Tonghua and Kang, Xuanwu and Wu, Xiaohan and Ding, Shijin and Zhang, David Wei},
 year = {2020},
 title = {Facile integration of MoS 2 /SiC photodetector by direct chemical vapor deposition},
 pages = {3035--3044},
 volume = {9},
 number = {9},
 issn = {2192-8614},
 journal = {Nanophotonics},
 doi = {\url{10.1515/nanoph-2019-0562}}
}

@article{Wei.2019,
 author = {Wei, Tingcha and Lau, Woon Ming and An, Xiaoqiang and Yu, Xuelian},
 year = {2019},
 title = {Interfacial Charge Transfer in MoS2/TiO2 Heterostructured Photocatalysts: The Impact of Crystal Facets and Defects},
 volume = {24},
 number = {9},
 journal = {Molecules (Basel, Switzerland)},
 doi = {\url{10.3390/molecules24091769}}
}

@article{Singh.2024,
 author = {Singh, Bheem and Gautam, Sudhanshu and Behera, Govinda Chandra and Kumar, Rahul and Aggarwal, Vishnu and Tawale, Jai Shankar and Ganesan, Ramakrishnan and Roy, Somnath Chanda and Kushvaha, Sunil Singh},
 year = {2024},
 title = {MoS 2 thin film decorated TiO 2 nanotube arrays on flexible Ti foil for solar water splitting application},
 pages = {015006},
 volume = {5},
 number = {1},
 journal = {Nano Express},
 doi = {\url{10.1088/2632-959X/ad1694}}
}

@article{Sim.2013,
 author = {Sim, Sangwan and Park, Jusang and Song, Jeong-Gyu and In, Chihun and Lee, Yun-Shik and Kim, Hyungjun and Choi, Hyunyong},
 year = {2013},
 title = {Exciton dynamics in atomically thin MoS 2 : Interexcitonic interaction and broadening kinetics},
 volume = {88},
 number = {7},
 issn = {1098-0121},
 journal = {Physical Review B},
 doi = {\url{10.1103/PhysRevB.88.075434}}
}

@article{Pollmann.2020b,
 author = {Pollmann, Erik and Madau{\ss}, Lukas and Schumacher, Simon and Kumar, Uttam and Heuvel, Flemming and {vom Ende}, Christina and Yilmaz, S{\"u}meyra and G{\"u}ng{\"o}rm{\"u}s, S{\"u}meyra and Schleberger, Marika},
 year = {2020},
 title = {Apparent differences between single layer molybdenum disulphide fabricated via chemical vapour deposition and exfoliation},
 pages = {505604},
 volume = {31},
 number = {50},
 journal = {Nanotechnology},
 doi = {\url{10.1088/1361-6528/abb5d2}}
}

@incollection{Pollmann.2018,
 author = {Pollmann, E. and Madau{\ss}, L. and Zeuner, V. and Schleberger, M.},
 title = {Strain in Single-Layer MoS2 Flakes Grown by Chemical Vapor Deposition},
 pages = {338--343},
 publisher = {Elsevier},
 isbn = {9780128098943},
 booktitle = {Encyclopedia of Interfacial Chemistry},
 year = {2018},
 doi = {\url{10.1016/B978-0-12-409547-2.14175-7}}
}

@article{Paul.2018,
 author = {Paul, Kamal Kumar and Mawlong, Larionette P. L. and Giri, P. K.},
 year = {2018},
 title = {Trion-Inhibited Strong Excitonic Emission and Broadband Giant Photoresponsivity from Chemical Vapor-Deposited Monolayer MoS2 Grown in Situ on TiO2 Nanostructure},
 pages = {42812--42825},
 volume = {10},
 number = {49},
 journal = {ACS applied materials {\&} interfaces},
 doi = {\url{10.1021/acsami.8b14092}}
}

@article{Liu.2019,
 author = {Liu, Huihui and Li, Yue and Xiang, Miaomiao and Zeng, Hualing and Shao, Xiang},
 year = {2019},
 title = {Single-Layered MoS2 Directly Grown on Rutile TiO2(110) for Enhanced Interfacial Charge Transfer},
 pages = {6083--6089},
 volume = {13},
 number = {5},
 journal = {ACS nano},
 doi = {\url{10.1021/acsnano.9b02608}}
}

@article{Kropp.2020,
 author = {Kropp, Jaron A. and Sharma, Ankit and Zhu, Wenjuan and Ataca, Can and Gougousi, Theodosia},
 year = {2020},
 title = {Surface Defect Engineering of MoS2 for Atomic Layer Deposition of TiO2 Films},
 pages = {48150--48160},
 volume = {12},
 number = {42},
 journal = {ACS applied materials {\&} interfaces},
 doi = {\url{10.1021/acsami.0c13095}}
}

@article{Huang.2021,
 author = {Huang, Chenxi and Fu, Jun and Xiang, Miaomiao and Zhang, Jiefu and Zeng, Hualing and Shao, Xiang},
 year = {2021},
 title = {Single-Layer MoS2 Grown on Atomically Flat SrTiO3 Single Crystal for Enhanced Trionic Luminescence},
 pages = {8610--8620},
 volume = {15},
 number = {5},
 journal = {ACS nano},
 doi = {\url{10.1021/acsnano.1c00482}}
}

@article{Haastrup.2023,
 author = {Haastrup, Mark J. and Bianchi, Marco and Lammich, Lutz and Lauritsen, Jeppe V.},
 year = {2023},
 title = {The interface ofin-situgrown single-layer epitaxial MoS2on SrTiO3(001) and (111)},
 volume = {35},
 number = {19},
 journal = {Journal of physics. Condensed matter : an Institute of Physics journal},
 doi = {\url{10.1088/1361-648X/acbf19}}
}

@article{Dong.2016,
 author = {Dong, X. and Yan, C. and Tomer, D. and Li, C. H. and Li, L.},
 year = {2016},
 title = {Spiral growth of few-layer MoS2 by chemical vapor deposition},
 volume = {109},
 number = {5},
 issn = {0003-6951},
 journal = {Applied Physics Letters},
 doi = {\url{10.1063/1.4960583}}
}

@inproceedings{Ding.2023,
 author = {Ding, Chengxi and Ma, Hongping},
 title = {Heteroepitaxial MoS 2 on Wide Bandgap Semiconductors: A Review},
 pages = {1--5},
 publisher = {IEEE},
 isbn = {979-8-3503-8537-3},
 booktitle = {2023 20th China International Forum on Solid State Lighting {\&} 2023 9th International Forum on Wide Bandgap Semiconductors (SSLCHINA: IFWS)},
 year = {2023},
 doi = {\url{10.1109/SSLChinaIFWS60785.2023.10399672}}
}

@article{Chen.2018,
 author = {Chen, Peiyu and Xu, Wenshuo and Gao, Yakun and Warner, Jamie H. and Castell, Martin R.},
 year = {2018},
 title = {Epitaxial Growth of Monolayer MoS 2 on SrTiO 3 Single Crystal Substrates for Applications in Nanoelectronics},
 pages = {6976--6988},
 volume = {1},
 number = {12},
 issn = {2574-0970},
 journal = {ACS Applied Nano Materials},
 doi = {\url{10.1021/acsanm.8b01792}}
}

@article{Zheng.2023,
 author = {Zheng, Peiming and Wei, Wenya and Liang, Zhihua and Qin, Biao and Tian, Jinpeng and Wang, Jinhuan and Qiao, Ruixi and Ren, Yunlong and Chen, Junting and Huang, Chen and Zhou, Xu and Zhang, Guangyu and Tang, Zhilie and Yu, Dapeng and Ding, Feng and Liu, Kaihui and Xu, Xiaozhi},
 year = {2023},
 title = {Universal epitaxy of non-centrosymmetric two-dimensional single-crystal metal dichalcogenides},
 pages = {592},
 volume = {14},
 number = {1},
 journal = {Nature communications},
 doi = {\url{10.1038/s41467-023-36286-6}}
}

@article{Park.2019,
 author = {Park, Taejin and Bae, Changdeuck and Lee, Hyangsook and Leem, Mirine and Kim, Hoijoon and Ahn, Wonsik and Kim, Jinbum and Lee, Eunha and Shin, Hyunjung and Kim, Hyoungsub},
 year = {2019},
 title = {Non-equilibrium fractal growth of MoS 2 for electrocatalytic hydrogen evolution},
 pages = {478--486},
 volume = {21},
 number = {3},
 journal = {CrystEngComm},
 doi = {\url{10.1039/C8CE01952E}}
}


\end{document}